\documentstyle[aps,prd]{revtex}
\begin{document}
\hfill{hep-ph/0203071}

\begin{center}
{\large {\bf $1/m_b$ source in inclusive $b$-hadron decays}}
\vskip 0.5cm
Huei-Shih Liao$^1$ and Hsiang-nan Li$^2$
\vskip 0.3cm
$^1$Department of Physics, National Tsing-Hua University,
Hsinchu, Taiwan 300, Republic of China
\vskip 0.3cm
$^2$Institute of Physics, Academia Sinica,
Taipei, Taiwan 115, Republic of China
\end{center}
\vskip 0.5cm

PACS numbers: 13.20.He, 12.38.Bx, 12.38.Cy, 14.40.Nd

\vskip 0.5cm
\centerline{\bf Abstract}

We show that threshold resummation of the end-point logarithms in
inclusive $b$-hadron decays leads to a Sudakov factor, which
introduces $1/m_b$ power corrections through infrared renormalons. These
power corrections are strong enough to account for the observed
semileptonic branching ratio, charm yield, and lifetime ratios by
breaking the local quark-hadron duality.

\vskip 0.5cm

Several puzzles in inclusive heavy hadron decays have existed for some
time. The observed semileptonic branching ratio $B_{\rm SL}$ and charm
yield $n_c$ of the inclusive $B$ meson decays were
$B_{\rm SL}=(10.19 \pm 0.37) \%$ and $n_c=1.12 \pm 0.05$ from CLEO
\cite{CLEO}, and $B_{\rm SL}=(11.12 \pm 0.20) \%$ and $n_c=1.20 \pm 0.07$
from LEP \cite{LEP}. For updated values, refer to \cite{ad}. The naive
parton model, which coincides with leading-power heavy quark effective
theory (HQET) \cite{CGG}, gives $B_{\rm SL} \sim 13\%$ \cite{a6}. The
$1/m_b^2$ corrections are less than $5\%$ of the leading ones
\cite{neubert}. Another puzzle is the low lifetime ratio
$\tau(\Lambda_b)/\tau(B_d)=0.794\pm 0.053$ \cite{ad}. The HQET prediction
up to $O(1/m_b^2)$ is about 0.99 \cite{NS}. When including the
$O(1/m_b^3)$ corrections, the ratio depends on six unknown parameters,
and reduces only to around 0.95 for various model estimates \cite{neubert}.

There has been the suggestion that $1/m_b$ corrections might appear,
{\it i.e.}, the local duality might break down in nonleptonic decays
\cite{grinstein}. It has been observed that if $m_b$ was replaced by the
$b$-hadon mass in the phase-space factor $m_b^5$ associated with
nonleptonic decays, the predictions agree well with the above data
\cite{A}. This modification reduces $B_{\rm SL}$ by increasing the
nonleptonic branching ratios, and explains the absolute $B$ meson decay
rate \cite{ae}. Note that HQET based on the $b$ quark kinematics accounts
for only 80\% of the decay rate. However, the mass replacement is lack
of a solid theoretical base.

In this letter we shall propose possible mechanism responsible for the
breakdown of the local duality, which arises from threshold resummation
associated with final-state particles. It has been shown that resummation
for the intial heavy hadron does not violate the duality \cite{LiYu}.
Define the variable $\bar\Lambda=m_B-m_b$ as the $B$ meson and $b$ quark
mass difference. In the kinematic region with the invariant mass of the
outgoing quark being $O(\bar\Lambda m_b)$, infrared divergences in the
decays are factorized into a $B$ meson structure function $f(k)$
\cite{N,BSV}, $k$ being the residual momentum carried by the $b$ quark.
The first moment of $f(k)$ vanishes, implying that
nonperturbative corrections start from $O(1/m_b^2)$.
In the end-point region with the invariant mass vanishing like
$O(\bar\Lambda^2)$, additional collinear divergences from loop momenta
parallel to the energetic quark momentum occur in higher-order
corrections, which demand the introduction of a jet function. Take the
$b\to s\gamma$ transition as an example. The end-point logarithms
$\alpha_s\ln(1-x)/(1-x)_+$ ($\alpha_s\ln^2 N$ in Mellin
space) with $x=2E_\gamma/m_b$, where $E_\gamma$ is the photon energy,
should be orgainzed into a jet function $J(1-x)$.

Threshold resummation of the above double logarithms leads to \cite{KS}
\begin{eqnarray}
J(1-x)=\int_{c-i\infty}^{c+i\infty}\frac{dN}{2\pi i}
{\tilde J}(N)x^{-N}\;,\;\;\;\;
{\tilde J}(N)=\exp\left[\int_0^{1} dz\frac{z^{N-1}-1}{1-z}
\int_{(1-z)^2}^{1-z}
\frac{d\lambda}{\lambda}\gamma_K(\alpha_s(\lambda m_b^2))\right]\;,
\label{mj}
\end{eqnarray}
where $c$ is an arbitrary real constant larger than the real parts of all
the poles in the Sudakov factor ${\tilde J}(N)$, and the anomalous
dimension is given, up to one loop, by $\gamma_K=\alpha_s C_F/\pi$,
$C_F=4/3$ being a color factor. Note that the physical range of $x$ is
$0\le x\le m_B/m_b$. We show that infrared renormalons in Eq.~(\ref{mj})
generate $1/m_b$ corrections by inserting the identity $\alpha_s(\mu^2)=\pi
\int_0^{\infty}d\sigma\exp[-\sigma \beta_1\ln(\mu^2/\Lambda^2_{\rm QCD})]$
into ${\tilde J}(N)$, with the coefficient $\beta_1=(33-2n_f)/12$, $n_f$
being the flavor number. Performing the integration over $\lambda$, we
have 
\begin{eqnarray}
{\tilde J}(N)=\exp\Bigg\{C_F\int_0^\infty \frac{d\sigma}{\sigma\beta_1}
\int_0^{1} dz\frac{z^{N-1}-1}{1-z}\left[
\frac{1}{(1-z)^{2\sigma\beta_1}}-\frac{1}{(1-z)^{\sigma\beta_1}}\right]
\left(\frac{\Lambda^2_{\rm QCD}}{m_b^2}\right)^{\sigma\beta_1}
\Bigg\}\;.
\label{jn}
\end{eqnarray}

Equation (\ref{jn}) can be expressed as a sum of contributions from all
the infrared renormalons. It is known that the leading infrared renormalon
from $\sigma\to 0$ corresponds to perturbative resummation. This can be
understood by considering the $\sigma\to 0$ limit of the integrand in
Eq.~(\ref{jn}):
\begin{eqnarray}
\exp\left[{\rm const.}\int_0^1dz\frac{z^{N-1}-1}{1-z}
\ln(1-z)\right]\equiv {\tilde J}(N)|_{\alpha_s}
\approx \exp\left(-\frac{1}{2}\gamma_K\ln^2 N\right)\;.
\label{jn1}
\end{eqnarray}
The integrand on the left-hand side has the same functional form as
in ${\tilde J}(N)$, if $\alpha_s$ was fixed, except the undetermined
const.. Hence, the left-hand side is equivalent to
${\tilde J}(N)|_{\alpha_s}$ at the leading-logarithm accuracy. Substitute
Eq.~(\ref{jn1}) into the inverse Mellin transformation in Eq.~(\ref{mj}),
and choose the branch cut along the negative real axis in the complex $N$
plane. For $0< x < 1$, we enclose the contour in Eq.~(\ref{mj}) toward
the minus real axis. The discontinuity crossing the branch cut gives,
using the variable change $w=\ln N$ \cite{L2},
\begin{eqnarray}
\lim_{x\to 1^-}J(1-x)|_{\alpha_s}=
-\exp\left(\frac{\gamma_K}{2}\pi^2+\frac{1}{2\gamma_K}\right)
\int_{-\infty}^{\infty}\frac{dw}{\pi} \sin (\pi\gamma_K w)
\exp\left(-\frac{1}{2}\gamma_K w^2\right)=0\;,
\end{eqnarray}
because the integrand is an odd function of $w$. For $x >1$, we enclose
the contour toward the positive real axis, where there is no cut and no
pole, indicating $J(1-x)|_{\alpha_s}=0$ for $x >1$. It is obvious from
Eq.~(\ref{mj}) that $J(1-x)|_{\alpha_s}$ vanishes at $x=0$. Therefore, we
obtain the qualitative behavior of $J(1-x)|_{\alpha_s}$: it takes a value
for $0<x<1$, and diminishes at $x=0$ and at $x=1$. 

The next-to-leading pole from $\sigma\to 1/(2\beta_1)$ leads to
\begin{eqnarray}
\exp\left[{\rm const.}
\int_0^1dz\frac{z^{N-1}-1}{1-z}\left(\frac{1}{1-z}
-\frac{1}{\sqrt{1-z}}\right)
\frac{\Lambda_{\rm QCD}}{m_b}\right]\;.
\label{j2}
\end{eqnarray}
We neglect the second term in the brackets, which is not important
compared to the first term, and employ the approximation
$z^{N-1}=(1+z-1)^{N-1}\approx 1+N(z-1)$, since the integral over $z$ is
dominated by the region $z\sim 1$. Equation (\ref{j2}) becomes
$\exp[{\rm const.}(\Lambda_{\rm QCD}/m_b)N]\approx
1 +{\rm const.}(\Lambda_{\rm QCD}/m_b)N$, up to corrections of
$O(\Lambda_{\rm QCD}^2/m_b^2)$, where the divergent integral
$\int dz/(1-z)$ has been absorbed into the ambigious const.. Picking up
the first two renormalons, we write
\begin{eqnarray}
{\tilde J}(N)={\tilde J}(N)|_{\alpha_s}
+{\rm const.}\frac{\Lambda_{\rm QCD}}{m_b}N
{\tilde J}(N)|_{\alpha_s}\;.
\end{eqnarray}

Inserting the above expression into Eq.~(\ref{mj}), the second term leads
to ${\rm const.}(\Lambda_{\rm QCD}/m_b)xJ'(1-x)|_{\alpha_s}$. To have 
meaningful predicitons, a $1/m_b$ power correction must appear to
cancel this ambigious term. Viewing its functional form, this $1/m_b$
correction can be introduced by Fourier expanding the jet function:
\begin{eqnarray}
J\left(1-t-\frac{\bar\Lambda}{m_b}t\right)
=J(1-t)-\frac{\bar\Lambda}{m_b}tJ'(1-t)\;,
\label{fex}
\end{eqnarray}
where the variable change $x=(m_B/m_b)t$, $0\le t\le 1$, has been made.
It can be shown that the $\sigma\to 1/(2\beta_1)$ renormalon in the
first term on the right-hand side of Eq.~(\ref{fex}) is cancelled
by the $\sigma\to 0$ renormalon in the second term under an appropriate
prescription for extracting the renormalon contribution. The details will
be published elsewhere. After this cancellation, we freeze the coupling
constants in the above two terms, and obtain well-defined predictions
for inclusive $b$-hadron decays.

We study the semileptonic decays $b\to c l\bar\nu$ for the lepton
$l=e,\mu,\tau$. Adopt the scaling variables $x=2E_l/m_B$ and
$y=q^2/m_B^2$ for the $b$ quark decays, where $E_l$ is the lepton energy
and $q \equiv p_l+p_{\nu}$ the momentum of the lepton pair. The physical
bounds in the ranges,
\begin{eqnarray}
2\alpha_l\leq & x &\leq 1+\alpha_l^2-\alpha_D^2\;,
\nonumber\\
\alpha_l^2+\left(1+\alpha_l^2-\alpha_D^2-x\right)
\frac{x-\sqrt{x^2-4\alpha_l^2}}{2-x+\sqrt{x^2-4\alpha_l^2}}
\leq &y&\leq \alpha_l^2+\left(1+\alpha_l^2-\alpha_D^2-x\right)
\frac{x+\sqrt{x^2-4\alpha_l^2}}{2-x-\sqrt{x^2-4\alpha_l^2}}\;,
\label{rkl1}
\end{eqnarray}
can be achieved, since the $b$ quark carries residual momentum \cite{N}.
The ratios are defined by $\alpha_{l,D,c}\equiv m_{l,D,c}/m_B$, $m_l$,
$m_D$ and $m_c$ being the lepton mass, the $D$ meson mass, and the charm
quark mass, respectively. $m_D$ appears as the minimal invariant mass of
the decay product $X_c$.

For these modes, loop corrections do not generate double logarithms,
because the $c$ quark is massive and collinear divergences are absent.
The factorization formula for the total decay width is then as simple as
\begin{equation}
\Gamma_{l\bar\nu}
= \frac{G_F^2|V_{cb}|^2}{8\pi^3}m^2_b\int dE_l dq^2
H\left(\frac{2E_l}{m_b},\frac{q^2}{m_b^2}\right)
=\Gamma_0\left(\frac{m_B}{m_b}\right)^3\int dxdy
H\left(\frac{m_B}{m_b}x,\frac{m_B^2}{m_b^2}y\right)\;,
\label{asb3}
\end{equation}
with $\Gamma_0 = G_F^2|V_{cb}|^2m^5_b/(16\pi^3)$ and the hard part
$H(x,y)=(1+y-x-m_c^2/m_b^2)(x-y-m_l^2/m_b^2)$. Equation (\ref{asb3})
does not involve the structure function $f(k)$, since the higher-power
contributions it introduces start from $O(1/m_b^2)$. Expanding the above
expression up to $\bar\Lambda/m_B$, we obtain
\begin{eqnarray}
\Gamma_{l\bar\nu}&=&\Gamma_0\int dxdy\bigg\{
\left(1+4\frac{\bar\Lambda}{m_b}\right)
(1+y-x-\alpha_c^2)(x-y-\alpha_l^2)
\nonumber\\
& &-\frac{\bar\Lambda}{m_B}
\left[(y+\alpha_l^2)(1-2x+3y-3\alpha_c^2)-x(2y-x-2\alpha_c^2)
\right]\bigg\}\;.
\label{asb4}
\end{eqnarray}

For the nonleptonic decays $b\to c\bar c s$, the threshold resummation
effect is associated with the energetic $s$ quark, and both the $c$ and
$\bar c$ quarks are regarded as being on-shell. The definitions of the
scaling variables are modified into $x=2E_c/m_B$,
$y=(p_c+p_{\bar c})^2/m_B^2$, and
$y_0=2(E_c+E_{\bar c})/m_B$, with $E_c$ ($E_{\bar c}$) and $p_c$
($p_{\bar c}$) being the $c$ ($\bar c$) quark energy and momentum,
respectively. Their physical ranges are
\begin{eqnarray}
2\alpha_c\leq &x& \leq 1
\nonumber\\
\frac{(1-x)(x-\sqrt{x^2-4\alpha_c^2})+2\alpha_c^2}
{2(1-x+\alpha_c^2)}
\leq &y&\leq
\frac{(1-x)(x+\sqrt{x^2-4\alpha_c^2})+2\alpha_c^2}
{2(1-x+\alpha_c^2)}\;,
\nonumber\\
\frac{1}{2\alpha_c^2}\left[xy-\sqrt{x^2-4\alpha_c^2}
\sqrt{y(y-4\alpha_c^2)}\right]
\leq &y_0& \le 1+y\;.
\label{rcs1}
\end{eqnarray}

The factorization formula for the total decay width is written as
\begin{equation}
\Gamma_{\bar c s}=\Gamma_0\left(\frac{m_B}{m_b}\right)^4\int dxdydy_0
c(m_b)H\left(\frac{m_B}{m_b}x,\frac{m_B^2}{m_b^2}y,
\frac{m_B}{m_b}y_0\right)
J\left(1-\frac{m_B}{m_b}y_0+\frac{m_B^2}{m_b^2}y\right)\;,
\label{acs1}
\end{equation}
with $H(x,y,y_0)=(y_0-x)(x-y)$. The factor $c(m_b)$ is given by
$c(m_b)=(N_c+1)c^2_+(m_b)/2+(N_c-1)c^2_-(m_b)/2$ with the color number
$N_c=3$ and $c_\pm=c_2\pm c_1$, where the Wilson coefficients $c_1$ and
$c_2$ correspond to the four-fermion operators
$O_1=({\bar s}_L\gamma_\mu b_L)({\bar c}_L\gamma^\mu c_L)$ and
$O_2=({\bar c}_L\gamma_\mu b_L)({\bar s}_L\gamma^\mu c_L)$, respectively.
We have confirmed that our predictions are insensitive to the 
choices of the arguments of the Wilson coefficients.
Expanding $H$ and $J$ up to $O(\bar\Lambda/m_b)$, Eq.~(\ref{acs1}) becomes
\begin{eqnarray}
\Gamma_{\bar c s}&=&\Gamma_0\int dxdydy_0 c(m_b)
\bigg\{\left(1+6\frac{\bar\Lambda}{m_b}\right)
(y_0-x)(x-y)\delta(1-y_0+y)
\nonumber\\
& &-\frac{\bar\Lambda}{m_b}\left[(y_0-x)yJ(1-y_0+y)
+(y_0-x)(x-y)(y_0-2y)J'(1-y_0+y)\right]\bigg\}\;.
\label{acs2}
\end{eqnarray}
Note that the jet function for the leading term, representing part of
perturbative corrections, has been set to the $\delta$-function. This is
for consistency, because the semileptonic decays are evaluated to leading
order in $\alpha_s$.

For the $b\to c \bar u d$ mode, we associate the threshold resummation
effect with the energetic $\bar u$ and $d$ quarks, and regard the $c$
quark as being on-shell. The scaling variables are defined by
$w=p_{\bar u}^2/m_B^2$, $x=2E_c/m_B$, $y=(p_c+ p_{\bar u})^2/m_B^2$, and
$y_0=2(E_c+E_{\bar u})/m_B$, whose ranges are 
\begin{eqnarray}
0\leq &w& \leq (1-\alpha_c)^2\;,
\nonumber\\
2\alpha_c\leq &x& \leq 1-w+\alpha_c^2\;,
\nonumber\\
\frac{x-\sqrt{x^2-4\alpha_c^2}}{2}
+\frac{2-x+\sqrt{x^2-4\alpha_c^2}}{2(1-x+\alpha_c^2)}w
\leq &y&\leq
\frac{x+\sqrt{x^2-4\alpha_c^2}}{2}
+\frac{2-x-\sqrt{x^2-4\alpha_c^2}}{2(1-x+\alpha_c^2)}w\;,
\nonumber\\
\frac{1}{2\alpha_c^2}\left[x(y-w+\alpha_c^2)
-\sqrt{x^2-4\alpha_c^2}\sqrt{(y-w-\alpha_c^2)^2-4\alpha_c^2 w}\right]
\le &y_0&\le 1+y\;,
\label{rud1}
\end{eqnarray}
The factorization formula for the total decay width is written,
up to $O(\bar\Lambda/m_b)$, as
\begin{eqnarray}
\Gamma_{\bar u d}
&=&\Gamma_0\int dwdxdydy_0 c(m_b)
\bigg\{\left(1+8\frac{\bar\Lambda}{m_b}\right)
(y_0-x)(w+x-y-\alpha_c^2)\delta(w)\delta(1-y_0+y)
\nonumber\\
& &-\frac{\bar\Lambda}{m_b}\left[
(y_0-x)(y-w+\alpha_c^2)J(w)J(1-y_0+y)
-2(y_0-x)(w+x-y-\alpha_c^2)wJ'(w)J(1-y_0+y)\right.
\nonumber\\
& &\left.+(y_0-x)(w+x-y-\alpha_c^2)(y_0-2y)J(w)J'(1-y_0+y)\right]\bigg\}\;.
\label{aud2}
\end{eqnarray}
The prefactors $\Gamma_0$ for the nonleptonic modes are the same as that
for the $b\to c\l\bar\nu$ decays with $|V_{ud}|=|V_{cs}|=1$.

Setting $m_B$ to $m_b$, {\it i.e.}, $\bar\Lambda=0$, Eqs.~(\ref{asb4}),
(\ref{acs2}) and (\ref{aud2}) reduce to the quark-level formalism.
To estimate the strength of the $\bar\Lambda/m_b$ corrections, we
propose the parametrization,
\begin{eqnarray}
J_p(1-t)&=&(a+1)(a+2)t^a(1-t)\;,
\label{ptr}
\end{eqnarray}
where the free parameter $a$ is determined by fitting ${\tilde J}_p(N)$
to ${\tilde J}(N)|_{\alpha_s}$ for the first few $N$. Equation (\ref{ptr})
is normalized to unity, approaches the $\delta$-function $\delta(1-t)$,
{\it i.e.}, the lowest-order expression of the jet function, as
$a\to \infty$ ($\alpha_s\to 0$), and obeys the boundary conditions
$J_p(0)=J_p(1)=0$. For a fixed $\alpha_s$, we have the moments
${\tilde J}(2)=\exp(-\gamma_K)$, ${\tilde J}(3)=\exp(-7\gamma_K/4)$,
$\cdots$. The fit of ${\tilde J}_p(N)$ to these moments implies
$a=15\sim 25$. We have confirmed that the numerical results are
insensitive to the variaton of $a$ within the above range, and to the
parametrization form of the jet function. Below we
shall choose $a=20$.

The above formalism applies to the $B_u$, $B_s$, and $\Lambda_b$ decays
simply by redefining the variable $\bar\Lambda=m_{H}-m_b$. It is observed
that the increase of the $b$ quark mass just decreases the total decay
rates, and has a small effect on the relative rates among different
modes. The predictions are more sensitive to the variation of the charm
mass: $B_{\rm SL}$ increases, while $n_c$ decreases with $m_c$. The total
decay rates also decrease with $m_c$. We
choose $m_b=4.8$ GeV (correspondng to $c(m_b)=3.24$) and $m_c=1.5$ GeV as
representative parameters. The results from the quark-level calculation
and from the inclusion of the $1/m_b$ corrections are displayed in Table
I, for which the masses $m_{B_d}=m_{B_u}=5.279$ GeV, $m_{B_s}=5.369$ GeV
and $m_{\Lambda_b}=5.621$ GeV have been adopted. It is found that the
$1/m_b$ corrections enhance the $b\to c\bar u d$ decay rate more than
the $b\to c\bar cs$ decay rate, such that
$B_{\rm SL}$ is reduced without increasing $n_c$.
The consistency with the observed
lifetime ratios $\tau(B_s)/\tau(B_d)$ and
$\tau(\Lambda_b)/\tau(B_d)$ is also improved. The ratio
$\tau(B_u)/\tau(B_d)$ remains to be unity, implying that the $B_d$ and
$B_u$ meson lifetime difference may be attributed to $1/m_b^2$ or
perturbative corrections \cite{CFL}.
Note that the absolute heavy hadron lifetimes for $|V_{cb}|=0.04$ are
still larger than the data by 30\%, thought the consistency is greatly
improved. This discrepancy is expected to be removed
by subleading contributions.

We thank A. Falk for useful discussion. This work was supported in part
by the National Science Council of R.O.C. under the Grant No.
NSC-90-2112-M-001-077 and by National Center for Theoretical Sciences,
R.O.C.

\vskip 0.5cm

Table I. The predicted
$B_{\rm SL}$, $n_c$, $\tau(B_u)/\tau(B_d)$, $\tau(B_s)/\tau(B_d)$, 
$\tau(\Lambda_b)/\tau(B_d)$, and $\tau(B_d)/\tau(B_d)_{\rm exp}$ for
$\tau(B_d)_{\rm exp}=1.56\times 10^{12}s$.
The data of the lifetime ratios are
$\tau(B_u)/\tau(B_d)=1.066\pm 0.020$ and
$\tau(B_s)/\tau(B_d)=0.945\pm 0.039$ \cite{ad}.

\vskip 0.5cm

\begin{center}
\begin{tabular}{ccccccc}
\hline
             & $B_{\rm SL}$ & $n_{c} $& $\tau(B_u)/\tau(B_d)$
& $\tau(B_s)/\tau(B_d)$  & $\tau(\Lambda_b)/\tau(B_d)$
& $\tau(B_d)/\tau(B_d)_{\rm exp}$ \\
\hline
quark level
&12.9\% & 1.18 & 1.0 & 1.0 & 1.0 & 2.33\\
$1/m_b$ correction
& 10.6\% & 1.19 & 1.0 & 0.93 & 0.75 & 1.32 \\
\hline
\end{tabular}
\end{center}

\end{document}